\documentclass[12pt, draftclsnofoot, onecolumn]{IEEEtran} %for IEEE Trans. on Comm

\usepackage{amsmath}
\usepackage{epsfig}
\usepackage{eucal}
\usepackage{amssymb}
\usepackage{bm}
\usepackage{cite}
\usepackage{graphicx}
\usepackage{multirow}
\usepackage{subfigure}
\usepackage{graphicx,color,psfrag}
\usepackage{multirow}

\usepackage{array, tabularx, colortbl, color, threeparttable}
\usepackage[table]{xcolor}
\usepackage{psfrag, pstricks}
\usepackage{booktabs}

\newtheorem{lem}{Lemma}
\newtheorem{thm}{Theorem}
\newtheorem{cor}{Corollary}
\newtheorem{Def}{Definition}

\newcommand{\SNR}{\mathsf{SNR}}

\newcommand{\PG}{\mathsf{PG}}

\newcommand{\paren}[1]{\left(#1\right)}
\newcommand{\sqparen}[1]{\left[#1\right]}
\newcommand{\brparen}[1]{\left\{#1\right\}}
\newcommand{\abs}[1]{\left| #1\right|}
\newcommand{\field}[1]{\ensuremath{\mathbb{#1}}}
\newcommand{\R}{\ensuremath{\field{R}}} % real numbers
\newcommand{\Rp}{\ensuremath{\R_+}} % positive real numbers
\newcommand{\I}[1]{\ensuremath{\mathsf{1}_{\left\{#1\right\}}}} % indicator function
\newcommand{\ra}{\ensuremath{\rightarrow}} % abbreviation for right arrow
\newcommand{\PR}[1]{\ensuremath{\mathsf{Pr}\left\{#1\right\}}} % probability with braces
\newcommand{\PRP}[1]{\ensuremath{\mathsf{Pr}\left(#1\right)}} %\probability with parentheses
\newcommand{\ES}[1]{\ensuremath{\mathsf{E}\left[#1 \right]}} %Expectation with square parentheses
\newcommand{\V}[1]{\ensuremath{\mathsf{Var}\left(#1 \right)}} %Variance with parentheses
\newcommand{\defeq}{\ensuremath{\triangleq}} %Triangle equation for definitions
\newcommand{\e}[1]{\ensuremath{{\rm e}^{#1}}} %Exponents of e

\newcommand{\sinr}{\ensuremath{{\sf SINR}}}
\newcommand{\snr}{\ensuremath{{\sf SNR}}}

\newcommand{\vecbold}[1]{\ensuremath{\boldsymbol{#1}}}

\renewcommand{\vec}[1]{\ensuremath{\boldsymbol{#1}}} %Re-define \vec command to generate vectors in bold

\newcommand\myatop[2]{\genfrac{}{}{0pt}{}{#1}{#2}}

\begin{document}
\title{Downlink Outage Performance of Heterogeneous Cellular Networks}

\author{Serkan Ak,~\IEEEmembership{Student Member,~IEEE}, Hazer Inaltekin,~\IEEEmembership{Member,~IEEE}, \\ H. Vincent Poor,~\IEEEmembership{Fellow,~IEEE} % <-this % stops a space
%\thanks{S. Ak and H. Inaltekin are with the Department of Electrical and Electronics Engineering, Antalya International University 07190, Antalya, Turkey (e-mail: \{serkan.ak, hazeri\}@antalya.edu.tr). H. Vincent Poor is with the Department of Electrical Engineering, Princeton University, Princeton, NJ 08544, USA (e-mail: poor@princeton.edu).

%This research was supported in part by a Marie Curie FP7-Reintegration-Grants within the 7th European Community Framework Programme under Grant PCIG10-GA-2011-303713, in part by the Scientific and Technological Research Council of Turkey (TUBITAK) under Grant 115E162, and in part by the U.S. National Science Foundation under Grant ECCS-1343210. %The material in this paper was presented in part at the Simons Conference on Networks and Stochastic Geometry, Austin, Texas, USA, May 2015. } 
} 

\markboth{}{Ak \MakeLowercase{\textit{et al.}}: Downlink Outage Performance of Heterogeneous Cellular Networks}

\maketitle
\begin{abstract}
This paper derives tight performance upper and lower bounds on the downlink outage efficiency of $K$-tier heterogeneous cellular networks (HCNs) for general signal propagation models with Poisson distributed base stations in each tier.  In particular, the proposed approach to analyze the outage metrics in a $K$-tier HCN allows for the use of general bounded path-loss functions and random fading processes of general distributions.  Considering two specific base station (BS) association policies, it is shown that the derived performance bounds track the actual outage metrics reasonably well for a wide range of BS densities, with the gap among them becoming negligibly small for denser HCN deployments. A simulation study is also performed for $2$-tier and $3$-tier HCN scenarios to illustrate the closeness of the derived bounds to the actual outage performance with various selections of the HCN parameters.        
\end{abstract}


%\begin{keywords}
%Cooperative transmission, channel estimation, power allocation.
%\end{keywords}

\IEEEpeerreviewmaketitle
\normalsize
\section{Introduction}\label{Introduction}
Fifth generation (5G) wireless networks are conceived as highly heterogeneous consisting of multiple-tiers of network elements with denser deployments of base stations (BSs) and more advanced communication protocols to deal with excessive data demand from mobile users \cite{Ghosh12, Andrews12, Hwang13, Hanzo16}.  Modeling and analyzing performance of such multi-tier heterogeneous cellular networks (HCNs) using spatial point processes have recently gained an increasing popularity \cite{Andrews11, Dhillon12, Jo12, Hossain13}. In particular, it is shown in \cite{Andrews11} that using a Poisson point process (PPP) model even for macro cell BS locations provides us with an approximation as good as the one provided by the conventional grid based model \cite{Macdonald79} for the actual network performance. %, with the PPP model leading to a lower performance bound and the grid model resulting in an upper performance bound. 
With the introduction of more irregularly deployed network elements such as femto BSs and distributed antennas, it is expected that the PPP model will more accurately track the actual HCN performance.  

A major design issue for such PPP based HCN models is to characterize the statistical properties of aggregate wireless interference (AWI) experienced by a mobile user by taking spatial positions of BSs and the random nature of the wireless channel into account.  This is possible for the special case of Rayleigh fading and the classical unbounded path-loss model to compute various performance metrics such as outage probability and ergodic rate in closed form\cite{Dhillon12, Jo12} but not so for general signal propagation models.  The aim of the present paper is to extend those previously known results and techniques to study the HCN performance to more general and heterogenous communication scenarios that incorporates general bounded path-loss models and general fading distributions.  This is achieved by leveraging the recent results approximating the standardized AWI distribution as a normal distribution \cite{Ak15}.    

In particular, we focus on the downlink outage performance in an HCN under two different BS association policies, named as the generic association policy and BARSS association policy.  We obtain tight performance bounds on the downlink outage probability and outage capacity in HCNs under both association policies for general signal propagation models.  The derived bounds approximate the actual HCN outage performance accurately for a wide range of BS densities in each tier.  Further, the gap between bounds become negligibly small as the BSs are more densely deployed.  Finally, the proposed approach can be extended to other association policies and performance metrics, with the potential of shedding light into HCN performance and design beyond specific selections of the path-loss model and the fading distribution.

\section{System Model}\label{System_Model}

In this section, we will introduce the details of the studied downlink model in a $K$-tier cellular topology, the signal propagation model and the association policy under which the outage performance of a HCN is determined.  

\subsection{The Downlink Model in a $K$-Tier Cellular Topology} 
We consider an overlay $K$-tier HCN in which the BSs in all tiers are fully-loaded (i.e., no empty queues) and access to the same communication resources both in time and frequency.  The BSs in different tiers are differentiated mainly on the basis of their transmission powers, with $P_k > 0$ being the transmission power of a tier-$k$ BS for $k = 1, \ldots, K$. As is standard in stochastic geometric modeling, it is assumed that BSs in tier-$k$ are distributed over the plane according to a homogeneous PPP $\Phi_k$ with intensity parameter $\lambda_k$ [BSs per unit area].  For the whole HCN, the aggregate BS location process, which is the superposition of all individual position processes, is denoted by $\Phi = \bigcup_{k = 1}^K \Phi_k$.

We place a test user at the origin and consider signals coming from all BSs, apart from the BS to which the test user associates itself, as the {\em downlink} AWI experienced by this test user.\footnote{Limiting our analysis to the test user located at the origin does not cause any loss of generality because the BS locations are determined according to homogeneous PPPs.} Since we focus on the downlink analysis, we assume that the uplink and downlink do not share any common communication resources.  Therefore, the uplink interference can be ignored for discovering the downlink outage performance experienced by the test user.

\subsection{The Signal Propagation Model}  
We model the large scale signal attenuation for tier-$k$, $k=1, \ldots, K$, by a {\em bounded}, monotone, non-increasing and continuous path-loss function $G_k: [0, \infty) \mapsto [0, \infty)$.  $G_k$ asymptotically decays to zero at least as fast as $t^{-\alpha_k}$ for some path-loss exponent $\alpha_k > 2$. The fading (power) coefficient for the wireless link between a BS located at point $\vecbold{X} \in \Phi$ and the test user is denoted by $H_{\vecbold{X}}$. The fading coefficients $\brparen{H_{\vecbold{X}}}_{\vecbold{X} \in \Phi}$ form a collection of independent random variables (also independent of $\Phi$), with those belonging to the same tier, say tier-$k$, having a common probability distribution function (PDF) $q_k(h), h \geq 0$.  The first, second and third order moments of fading coefficients are assumed to be finite, and are denoted by $m^{(k)}_{H}$, $m^{(k)}_{H^2}$ and $m^{(k)}_{H^3}$, respectively, for tier-$k$.  We note that this signal propagation model is general enough that $H_{\vecbold{X}}$'s could also be thought to incorporate {\em shadow fading} effects due to blocking of signals by large obstacles existing in the communication environment, although we do not model such random factors explicitly and separately in this paper.

\subsection{Association Policy and Interference Power}
Association policy is a key mechanism that determines the outage performance experienced by the test user as it regulates the useful signal power as well as the interference power at the test user. Hence, we first formally define it to facilitate the upcoming discussion. 

\begin{Def}
An association policy $\mathcal{A}: \Omega \times \Rp^\infty \times \Rp^K \times \Rp^K \mapsto \R^2$ is a mapping that takes a BS configuration $\varphi \in \Omega$ (i.e., a countable point measure), fading coefficients $\brparen{h_{\vecbold{x}}}_{\vecbold{x} \in \varphi}$, transmission power levels $\brparen{P_k}_{k=1}^K$ and biasing factors $\brparen{\beta_k}_{k=1}^K$ as an input and determines the BS location to which the test user is associated as an output.   
\end{Def}

For the HCN model explained above, the output of $\mathcal{A}$ is a random point $\vecbold{X}^\star  = \paren{X_1^\star, X_2^\star} \in \Phi$ since the BS locations and fading coefficients are random elements. Biasing coefficients are important design parameters to offload data from bigger cells to the smaller ones.  Two other important random quantities related to $\vecbold{X}^\star$ is the tier index $A^\star$ to which $\vecbold{X}^\star$ belongs and the distance between $\vecbold{X}^\star$ and the origin, which is denoted by $R^\star = \left\| \vecbold{X}^\star \right\|_2 = \sqrt{\paren{X_1^\star}^2 + \paren{X_2^\star}^2}$.  Using these definitions, the total interference power at the test user is written as
$$ I_{\vecbold{\lambda}} = \sum_{\vecbold{X} \in \Phi \backslash \brparen{\vecbold{X}^\star}} P_{\vecbold{X}} H_{\vecbold{X}} G_{\vecbold{X}}\paren{\left\| \vecbold{X} \right\|_2}, $$   
 where $\vecbold{\lambda} = \sqparen{\lambda_1, \ldots, \lambda_K}^\top$, and it is understood that $P_{\vecbold{X}} = P_k$ and $G_{\vecbold{X}} = G_k$ if $\vecbold{X} \in \Phi_k$. 
 
The signal-to-interference-plus-noise ratio (\sinr) is the main performance determinant for the HCN model in question. Given an association policy $\mathcal{A}$, the $\sinr$ level experienced by the test user is equal to 
$$ \sinr_{\mathcal{A}} = \frac{P_{A^\star} H_{\vecbold{X}^\star} G_{A^\star}\paren{R^\star}}{N_0 + \frac{1}{\PG} I_{\vecbold{\lambda}}},$$
where $N_0$ is the constant background noise power and $\PG \geq 1$ is the processing gain constant that signifies the interference reduction capability, if possible, of the test user.  We also let $\SNR_k = \frac{P_k}{N_0}$ to denote the signal-to-noise ratio ($\SNR$) for tier-$k$.  Next, we define the main performance metrics used to measure the HCN outage performance.  
\begin{Def} \label{Def: Outage Metrics}
For a target bit rate $\tau$, $\tau$-outage probability is equal to 
$$\PRP{\tau\mbox{-outage}} = \PR{\log\paren{1 + \sinr_{\mathcal{A}}} < \tau}.$$ 
Similarly, for a target outage probability $\gamma$, the outage capacity is equal to
$$ C_{\rm o}\paren{\gamma} = \sup\brparen{\tau \geq 0: \PRP{\tau\mbox{-outage}} \leq \gamma},$$
which is the maximum data rate supported with outage probability not exceeding $\gamma$.   
\end{Def}

In the next section, we will derive the performance bounds on the HCN outage metrics above under two specific association policies: (i) a generic association policy and (ii) biased average received signal strength (BARSS) association policy. However, it should be noted that the analytical approach developed below is general enough for any association policy that preserves Poisson distribution property for BS locations given the information of $\vecbold{X}^\star$, with conditional {\em non-homogeneous} BS distributions allowed under such information.        

\section{Bounds on the HCN Outage Performance} \label{Section: HCN Outage Performance}

In this section, we introduce two specific association policies and derive the bounds on the HCN outage performance for these association policies. The long proofs are relegated to the appendices. Hence, we focus on the main engineering and design implications of these results for emerging 5G networks in the main body of the paper.    

\subsection{Generic Association Policy} 

We start our discussion with the generic association policy. The generic association policy is the policy under which the test user is connected to a BS in tier-$k$ at a distance $r$ from the origin, and the locations of the rest of the (interfering) BSs in each tier form a homogeneous PPP over $\R^2\backslash \mathcal{B}\paren{\vecbold{0}, d_i}$ given this connection information for $i=1, \ldots, K$, where $\mathcal{B}\paren{\vecbold{0}, d_i}$ is the {\em planar} ball centered at the origin with radius $d_i \geq 0$. $\mathcal{B}\paren{\vecbold{0}, d_i}$ can be thought to signify an {\em exclusion} region around the test user due to operation of the HCN network protocol stack.   

The study of the generic association policy, which may seem a little artificial at the first sight, will set the stage for us to analyze the outage performance of the BARSS association policy in the next part of the paper. The following lemma is obtained by specializing Theorem 1 from \cite{Ak15} to this case. It establishes the Gaussian approximation bounds for the distribution of (standardized) $I_{\vecbold{\lambda}}$.    
\begin{lem} \label{Lemma: Gauss Approximation for Generic Policy}
Under the generic association policy described above, for all $x \in \R$,
$$\abs{\PR{\widehat{I}_{\vecbold{\lambda}} \leq x} - \Psi(x)} \leq \Xi \cdot c(x), $$
where $\widehat{I}_{\vecbold{\lambda}} = \frac{I_{\vecbold{\lambda}} - \ES{I_{\vecbold{\lambda}}}}{\sqrt{\V{I_{\vecbold{\lambda}}}}}$, $c(x) = \min\paren{0.4785, \frac{31.935}{1+\abs{x}^3}}$, $\Xi = \frac{1}{\sqrt{2\pi}} \sum_{i=1}^K \frac{\lambda_i P_i^3 m_{H^3}^{(i)} \int_{d_i}^\infty G_i^3(t) t dt}{\paren{\sum_{i=1}^K \lambda_i P_i^2 m_{H^2}^{(i)} \int_{d_i}^\infty G_i^2(t) t dt}^\frac32}$ and $\Psi(x) = \frac{1}{\sqrt{2\pi}}\int_{-\infty}^x \e{- \frac{t^2}{2}} dt$, which is the standard normal cumulative distribution function (CDF).   
\end{lem}

Since the outage metrics given in Definition \ref{Def: Outage Metrics} heavily depend on the level of AWI at the test user, the above Gaussian approximation bounds play a key role to obtain the following upper and lower bounds on the outage probability. 
\begin{thm} \label{Thm: Outage Probability Bounds - Generic Association}
Let $\zeta_k\paren{h, \tau, r} =\frac{P_k \paren{\frac{h G_k(r)}{\e{\tau} - 1} - \SNR^{-1}_k}\PG - \ES{I_{\vecbold{\lambda}}}}{\sqrt{\V{I_{\vecbold{\lambda}}}}}$.  Then, $\PRP{\tau\mbox{-outage}}$ under the generic association policy is bounded above and below as
$$1 - \ES{V_k^+\paren{H_k, \tau, r}} \leq \PRP{\tau\mbox{-outage}} \leq 1 - \ES{V_k^-\paren{H_k, \tau, r}},$$
where $H_k$ is a generic random variable with PDF $q_k$, and the functions $V_k^+$ and $V_k^-$ are given as %in \eqref{Eqn: Vk plus} and \eqref{Eqn: Vk minus}, respectively.
\begin{equation}
V_k^+\paren{h, \tau, r} = \min\brparen{1, \Psi\paren{\zeta_k\paren{h, \tau, r}} + \Xi \cdot c\paren{\zeta_k\paren{h, \tau, r}}} \I{h \geq \frac{\SNR_k^{-1}\paren{\e{\tau} -1}}{G_k(r)}} \label{Eqn: Vk plus}
\end{equation}
and
\begin{equation}
V_k^-\paren{h, \tau, r} = \max\brparen{0, \Psi\paren{\zeta_k\paren{h, \tau, r}} - \Xi \cdot c\paren{\zeta_k\paren{h, \tau, r}}} \I{h \geq \frac{\SNR_k^{-1}\paren{\e{\tau} -1}}{G_k(r)}}. \label{Eqn: Vk minus}
\end{equation}    
\end{thm}
\begin{IEEEproof}
Please see Appendix \ref{Proof_Theorem_1}.
\end{IEEEproof} 

%%\small
%\begin{figure*}[!t]
%\begin{equation}
%V_k^+\paren{h, \tau, r} = \min\brparen{1, \Psi\paren{\zeta_k\paren{h, \tau, r}} + \Xi \cdot c\paren{\zeta_k\paren{h, \tau, r}}} \I{h \geq \frac{\SNR_k^{-1}\paren{\e{\tau} -1}}{G_k(r)}} \label{Eqn: Vk plus}
%\end{equation}
%\hrulefill
%\begin{equation}
%V_k^-\paren{h, \tau, r} = \max\brparen{0, \Psi\paren{\zeta_k\paren{h, \tau, r}} - \Xi \cdot c\paren{\zeta_k\paren{h, \tau, r}}} \I{h \geq \frac{\SNR_k^{-1}\paren{\e{\tau} -1}}{G_k(r)}} \label{Eqn: Vk minus}
%\end{equation} 
%\hrulefill
% The spacer can be tweaked to stop underfull vboxes.
%\vspace*{0pt}
%\end{figure*}
%%\normalsize

Using the bounds on $\PRP{\tau\mbox{-outage}}$, we can also bound $C_{\rm o}\paren{\gamma}$ for the generic association policy as below.
\begin{thm} \label{Thm: Outage Capacity Generic Association}
$C_{\rm o}\paren{\gamma}$ under the generic association policy is bounded above and below as
$$ C_{\rm o}\paren{\gamma} \leq \sup\brparen{\tau \geq 0: 1 - \ES{V_k^+\paren{H_k, \tau, r}} \leq \gamma}$$
and
$$ C_{\rm o}\paren{\gamma} \geq \sup\brparen{\tau \geq 0: 1 - \ES{V_k^-\paren{H_k, \tau, r}} \leq \gamma}. $$
\end{thm}
\begin{IEEEproof}
The proof follows from observing that the upper (lower) bound on the outage probability crosses the target outage probability $\gamma$ earlier (later) than $\PRP{\tau\mbox{-outage}}$ as $\tau$ increases.  
\end{IEEEproof}  

An important high level perspective about the detrimental effects of the network interference on the HCN outage performance can be obtained if we study the outage capacity bounds given in Theorem \ref{Thm: Outage Capacity Generic Association} as a function of $\vecbold{\lambda}$.  At each fading state $H_k = h$, it can be shown that the outage capacity scales with the BS intensity parameters according to $\Theta\paren{\frac{1}{\| \vecbold{\lambda} \|_2}}$ as $\| \vecbold{\lambda} \|_2$ grows to infinity.\footnote{$f\left( t \right)$ is said to be $\Theta \left( {g\left( t \right)} \right)$  as $t \ra t_0$ if $\mathop {\lim \sup }\limits_{t \to {t_0}} \frac{{f\left( t \right)}}{{g\left( t \right)}} < \infty $ and $\mathop {\lim \inf }\limits_{t \to {t_0}} \frac{{f\left( t \right)}}{{g\left( t \right)}} > 0$.}  This observation is different than the scale-invariance property of $\sinr$ statistics with BS density observed in some previous work such as \cite{Dhillon12} and \cite{Jo12}.  The main reason is that the increase in $I_{\vecbold{\lambda}}$ with denser HCN deployments cannot be counterbalanced by an increase in the received power levels for bounded path-loss models.  From an HCN design perspective, this result implies that it is imperative to set BS intensities at each tier appropriately for the proper deliver of data services with minimum required QoS to the end users.     

\subsection{BARSS Association Policy}
Now, we study the HCN outage performance under the BARSS association policy, in which the test user associates itself to the BS $\vecbold{X}^\star$ given by
$$ \vecbold{X}^\star = \mathop{\arg\max}_{\vecbold{X} \in \Phi} \beta_{\vecbold{X}} P_{\vecbold{X}} G_{\vecbold{X}}\paren{\| \vecbold{X} \|_2}, $$ 
where it is understood that $\beta_{\vecbold{X}} = \beta_k$ if $\vecbold{X} \in \Phi_k$. Consider the event $E_k(r)$ that $A^\star = k$ and $R^\star = r$, i.e., $E_k(r)$ is the event that the test user is associated to a tier-$k$ BS at a distance $r$ under the BARSS association policy.  It is easy to see that the locations of BSs in tier-$i$ form a homogeneous PPP over $\R^2 \backslash \mathcal{B}\paren{\vecbold{0}, Q_i^{(k)}(r)}$ given the event $E_k(r)$ for $i=1, \ldots, K$, where $Q_i^{(k)}(r) = G_i^{-1}\paren{\frac{\beta_k P_k}{\beta_i P_i} G_k(r)}$, and $G_i^{-1}(y) = \inf\brparen{x \geq 0: G_i(x) = y}$ if $y \in \sqparen{0, G_i(0)}$ and zero otherwise. %\footnote{It must be noted that we will have a different definition of $Q_i$ depending on the tier index $k$ we focus on, i.e., $Q_1$ can be a different function for $k=2$ and $k=3$, but we do not show this dependence explicitly to simplify the notation.}  
This observation puts us back to the generic association policy framework, and the derivation of the bounds for the conditional outage probability/capacity on the conditioned event $E_k(r)$ proceeds as before. Averaging over the event $E_k(r)$, we obtain the bounds for the unconditional outage probability and capacity metrics. To this end, we need the following lemmas. 
\begin{lem} \label{Lemma: Gauss Approximation for BARSS Policy}
Under the BARSS association policy described above, for all $x \in \R$,
$$\abs{\PR{\widehat{I}_{\vecbold{\lambda}} \leq x \ \big| \ E_k(r)} - \Psi(x)} \leq \Xi_k\paren{r} \cdot c(x), $$
where $\Xi_k(r) = \frac{1}{\sqrt{2\pi}} \sum_{i=1}^K \frac{\lambda_i P_i^3 m_{H^3}^{(i)} \int_{Q_i^{(k)}(r)}^\infty G_i^3(t) t dt}{\paren{\sum_{i=1}^K \lambda_i P_i^2 m_{H^2}^{(i)} \int_{Q_i^{(k)}(r)}^\infty G_i^2(t) t dt}^\frac32}$, and $\widehat{I}_{\vecbold{\lambda}}$, $c(x)$ and $\Psi(x)$ are as defined in Lemma \ref{Lemma: Gauss Approximation for Generic Policy}. 
\end{lem} 

We note that this is almost the same result appeared in Lemma \ref{Lemma: Gauss Approximation for Generic Policy}, except a small change in the definition of the constant $\Xi$ to show its dependence on the conditioned event $E_k(r)$.  In order to achieve averaging over the event $E_k(r)$, we need to know the connection probability to a tier-$k$ BS and the conditional PDF of the connection distance given that the test user is associated with a tier-$k$ BS.  To this end, we first obtain the connection probability to a tier-$k$ BS, which we denote by $p_k^\star \defeq \PR{A^\star = k}$, in the next lemma.       
\begin{lem} \label{Lemma: Association Probability}
Let $a_0 = 0$, $a_{K+1} = +\infty$ and $a_i = \frac{\beta_i P_i}{\beta_k P_k} G_i(0)$ for $i \in \brparen{1, \ldots, K} \backslash \brparen{k}$. Let $\pi(i)$ be an enumeration of $a_i$'s in descending order, i.e., $a_{\pi(i)} \geq a_{\pi(i+1)}$ for $i=0, \ldots, K-1$. Let $r_i = G_k^{-1}\paren{a_{\pi(i)}}$ for $i=0, \ldots, K$. Then, $p_k^\star$ is given by 
\begin{equation}
p_k^\star = 2\pi\lambda_k \sum_{j=1}^{K} \int_{r_{j-1}}^{r_j} u \exp\paren{-\pi \paren{\lambda_k u^2 + \sum_{i=1}^{j-1}\lambda_{\pi(i)} \paren{Q_{\pi(i)}^{(k)}\paren{u}}^2}}du. \label{Eqn: Association Probability} 
\end{equation}
%by \eqref{Eqn: Association Probability}. 
\end{lem}
\begin{IEEEproof}
Please see Appendix \ref{Proof_Lemma_3}.
\end{IEEEproof}

Several important remarks are in order regarding Lemma \ref{Lemma: Association Probability}.  The integration in \eqref{Eqn: Association Probability} is with respect to the nearest neighbour distance distribution for tier-$k$ to which the test user is associated. Hence, the BSs in some tiers are inactive to contribute to the association probability for different ranges of the nearest distance from $\Phi_k$ to the origin. This behaviour is different than that observed in \cite{Jo12}, which is again a manifestation of the bounded nature of the path-loss model.  In the next lemma, we derive the PDF of $R^\star$ given $A^\star = k$.        
\begin{lem} \label{Conditional_Connection_Distance}
Let $a_0 = 0$, $a_{K+1} = +\infty$ and $a_i = \frac{\beta_i P_i}{\beta_k P_k} G_i(0)$ for $i \in \brparen{1, \ldots, K} \backslash \brparen{k}$. Let $\pi(i)$ be an enumeration of $a_i$'s in descending order, i.e., $a_{\pi(i)} \geq a_{\pi(i+1)}$ for $i=0, \ldots, K-1$. Let $r_i = G_k^{-1}\paren{a_{\pi(i)}}$ for $i=0, \ldots, K$.  Then, the conditional PDF $f_k(u)$ of $R^\star$ given $A^\star = k$ is given as 
\begin{equation}
f_k(u) = \frac{2\pi\lambda_k}{p_k^\star} \sum_{j=1}^{K} u \exp\paren{-\pi\paren{\lambda_k u^2 + \sum_{i=1}^{j-1} \lambda_{\pi(i)}\paren{Q_{\pi(i)}^{(k)}(u)}^2}} \I{u \in [r_{j-1}, r_j)}. \label{Eqn: Conditional Connection Distance}
\end{equation}
%by \eqref{Eqn: Conditional Connection Distance}. 
\end{lem}
\begin{IEEEproof}
Please see Appendix \ref{proof_Conditional_Connection_Distance}.
\end{IEEEproof} 

The conditional connection PDF $f_k(u)$ given in \eqref{Eqn: Conditional Connection Distance} can be simplified significantly for small number of tiers.  A reduced expression for one particular but important case of a two-tier HCN is given by the following corollary.  
\begin{cor}\label{Corollary: Two-Tier Connection Distance}
Assume $K=2$, $\beta_1 P_1 G_1(0) \leq \beta_2 P_2 G_2(0)$ and $u^\star = G_2^{-1}\paren{\frac{\beta_1 P_1}{\beta_2 P_2} G_1(0)}$. Then,
$$f_1(u) = \frac{2 \pi \lambda_1}{p_1^\star} u \exp\paren{-\pi\paren{\lambda_1 u^2 + \lambda_2 \paren{Q_2^{(1)}(u)}^2}} \I{u \geq 0} $$
and 
\begin{eqnarray*}
f_2(u) =  \frac{2 \pi \lambda_2}{p_2^\star} u\exp\paren{-\pi \lambda_2 u^2} \I{u < u^\star} + \frac{2 \pi \lambda_2}{p_2^\star} u \exp\paren{-\pi\paren{\lambda_2 u^2 + \lambda_1 \paren{Q_1^{(2)}(u)}^2}} \I{u \geq u^\star}.
%f_2(u) =  \frac{2 \pi \lambda_2}{p_2^\star} u\exp\paren{-\pi \lambda_2 u^2} \I{u < u^\star} \hspace{5cm} \\ \hspace{0cm} + \frac{2 \pi \lambda_2}{p_2^\star} u \exp\paren{-\pi\paren{\lambda_2 u^2 + \lambda_1 Q_1(u)}} \I{u \geq u^\star}. \hspace{2cm}
\end{eqnarray*}
\end{cor} 

%\small
%\begin{figure*}[!t]
%\begin{equation}
%p_k^\star = 2\pi\lambda_k \sum_{j=1}^{K} \int_{r_{j-1}}^{r_j} u \exp\paren{-\pi \paren{\lambda_k u^2 + \sum_{i=1}^{j-1}\lambda_{\pi(i)} Q_{\pi(i)}\paren{u}}}du \label{Eqn: Association Probability} 
%\end{equation}
%\hrulefill
%\begin{equation}
%f_k(u) = \frac{2\pi\lambda_k}{p_k^\star} \sum_{j=1}^{K} u \exp\paren{-\pi\paren{\lambda_k u^2 + \sum_{i=1}^{j-1} \lambda_{\pi(i)}Q_{\pi(i)}\paren{u}}} \I{u \in [r_{j-1}, r_j)} \label{Eqn: Conditional Connection Distance}
%\end{equation}
%\hrulefill
%% The spacer can be tweaked to stop underfull vboxes.
%\vspace*{0pt}
%\end{figure*}
%\normalsize

\begin{figure*}[!t]
\begin{minipage}[b]{0.45\linewidth} % A minipage that covers 0.45 of the page
\centering
\includegraphics[width=3.2in]{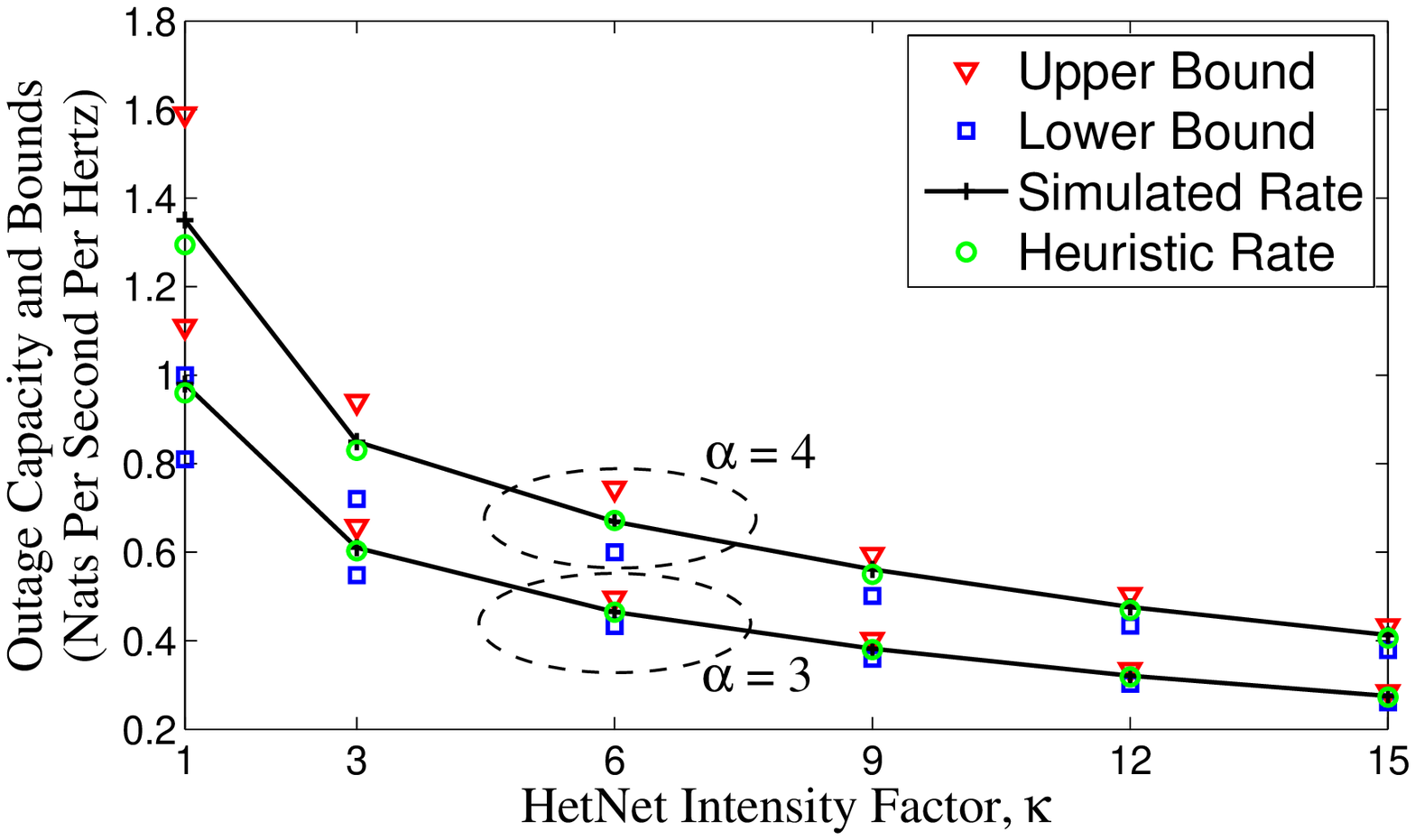}
\end{minipage}
\hspace{0.9cm} % To get a little bit of space between the figures
\begin{minipage}[b]{0.45\linewidth}
\centering
\includegraphics[width=3.2in]{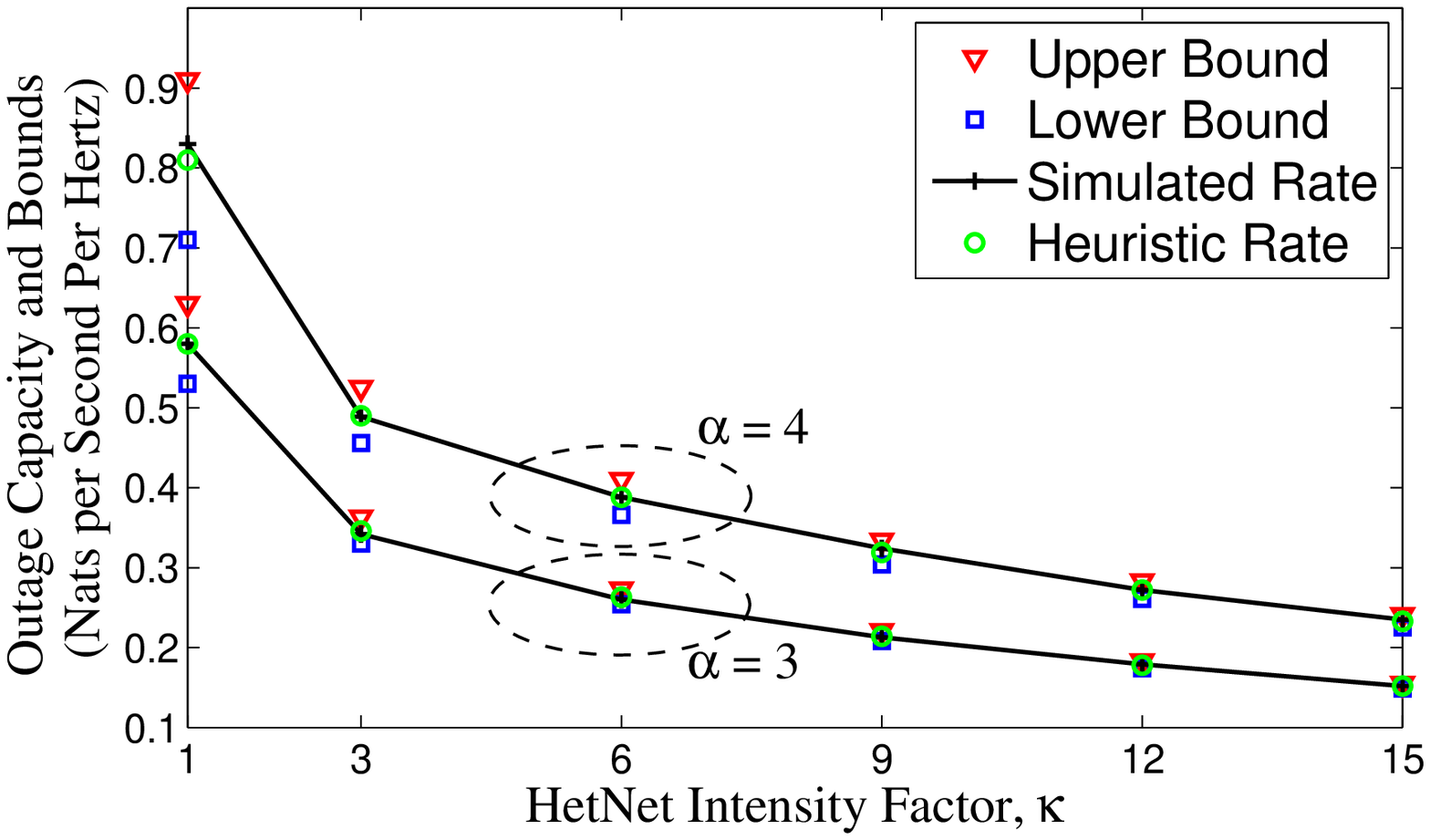}
\end{minipage}
\caption{Upper and lower bounds on ${C_{\rm o}}\left( \gamma  \right)$ for a 2- and 3-tier HCNs in the lefthand side and righthand side figures, respectively.}
\label{fig1}
\end{figure*}

Using these preliminary results, the performance bounds on the outage probability and capacity under the BARSS association policy are given in theorems below. 
\begin{thm} \label{Thm: Outage Probability - BARSS}
Let $\widehat{V}_k^{\pm}\paren{h, \tau, r}$ be defined as in \eqref{Eqn: Vk plus} and \eqref{Eqn: Vk minus}, respectively, by replacing $\Xi$ by $\Xi_k(r)$. Then, $\PR{\tau\mbox{-outage}}$ under the BARSS association policy is bounded below and above as
$$\PR{\tau\mbox{-outage}} \geq 1 - \sum_{k=1}^K p_k^\star \int_0^\infty f_k(r) \ES{\widehat{V}_k^+\paren{H_k, \tau, r}} dr$$
and
$$\PR{\tau\mbox{-outage}} \leq 1 - \sum_{k=1}^K p_k^\star \int_0^\infty f_k(r) \ES{\widehat{V}_k^-\paren{H_k, \tau, r}} dr.$$
\end{thm}
\begin{IEEEproof}
The proof follows from calculating these bounds for $\PR{\tau\mbox{-outage} \ | \ E_k(r)}$ using Theorem \ref{Thm: Outage Probability Bounds - Generic Association}, and then averaging them by using \eqref{Eqn: Association Probability} and \eqref{Eqn: Conditional Connection Distance}.  
\end{IEEEproof}
\begin{thm} \label{Thm: Outage Capacity - BARSS}
$C_{\rm o}\paren{\gamma}$ under the BARSS association policy is bounded above and below as
%\footnotesize
$$ C_{\rm o}\paren{\gamma} \leq \sup\brparen{\tau \geq 0: 1 - \sum_{k=1}^K p_k^\star \int_0^\infty f_k(r) \ES{\widehat{V}_k^+\paren{H_k, \tau, r}} dr \leq \gamma}$$
\normalsize
and
%\footnotesize
$$ C_{\rm o}\paren{\gamma} \geq \sup\brparen{\tau \geq 0: 1 - \sum_{k=1}^K p_k^\star \int_0^\infty f_k(r) \ES{\widehat{V}_k^-\paren{H_k, \tau, r}} dr \leq \gamma}.$$
%\normalsize
\end{thm}
\begin{IEEEproof}
The proof follows from observing that the upper (lower) bound on the outage probability
crosses the target outage probability $\gamma$ earlier (later) than $\PRP{\tau\mbox{-outage}}$ as $\tau$ increases.  
\end{IEEEproof}  
\vspace{-0.3cm}
\section{Simulation Results}\label{Simulation_Results}
In this part, we present our simulation results illustrating the upper and lower bounds on the HCN outage metrics derived in Section \ref{Section: HCN Outage Performance}.  In particular, we will only investigate $C_{\rm o}\paren{\gamma}$ for $2$- and $3$-tier HCNs under the BARSS association policy with no biasing. $N_0$ is set to zero and all fading coefficients are independently drawn from Nakagami-$m$ distribution with unit mean power gain and $m = 5$. The path-loss function is taken to be $G(x) = \frac{1}{1 + x^\alpha}$ for all tiers. The transmission powers are set as $P_1 = 4 P_2 = 16 P_3$, while we set BS intensities as $\lambda_1 = 0.1 \kappa$, $\lambda_2 = \kappa$ and $\lambda_3 = 5 \kappa$. Here, $\kappa$ is our control parameter to control the average number of BSs per unit area.  For the $2$-tier scenario, only $\brparen{P_k, \lambda_k}_{k=1}^2$ are considered.  The target outage probability is $0.15$ for Fig. \ref{fig1} and $\PG$ is set to $25$ for both figures. 

We plot the bounds in Theorem \ref{Thm: Outage Capacity - BARSS} on ${C_{\rm o}}\left( \gamma  \right)$ for $2$- and $3$-tier HCNs as a function of $\kappa$ in Fig. \ref{fig1}. Two different values of $\alpha$ are used.  As this figure shows, both upper and lower bounds approximate $C_{\rm o}\paren{\gamma}$ within $0.06$ Nats$/$Sec$/$Hz for $\alpha=3$ and within $0.15$ Nats$/$Sec$/$Hz for $\alpha = 4$ in the $2$-tier scenario.  They are tighter for the $3$-tier scenario due to denser HCN deployment. The heuristic rate curve, which is the average of the upper and lower bounds, almost perfectly track $C_{\rm o}\paren{\gamma}$ for all cases considered in Fig. \ref{fig1}.       

An interesting observation is the monotonically decreasing nature of $C_{\rm o}\paren{\gamma}$ with $\kappa$. This is in accordance with the discussion on the $\Theta\paren{\frac{1}{\| \vecbold{\lambda} \|_2}}$-type scaling behaviour of outage capacity in Section \ref{Section: HCN Outage Performance}.  Hence, we cannot improve the downlink data rates indefinitely in an HCN by adding more BS infrastructure.  We must either mitigate interference more efficiently or find the optimum BS density per tier maximizing delivered data rates per unit area.

\begin{figure}[!t]
\centering
\includegraphics[width=3.5in]{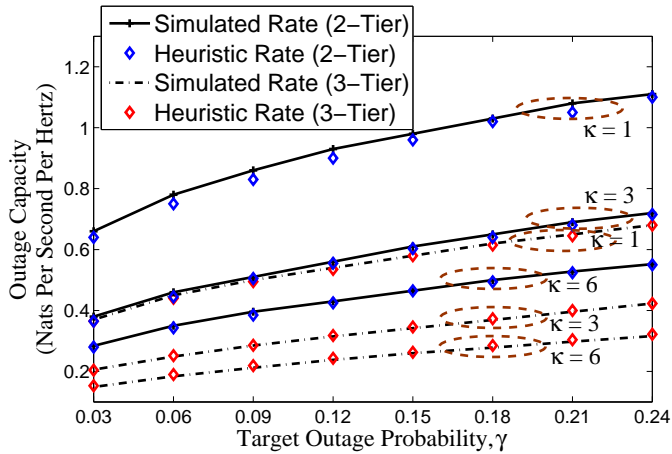}
\caption{Change of ${C_{\rm o}}\paren{\gamma}$ as a function of $\gamma$ for various values of $\kappa$. $\alpha$ is set to $3$. }
\label{fig2}
\end{figure}

Finally, we demonstrate $C_{\rm o}\paren{\gamma}$ as a function of $\gamma$ in Fig. \ref{fig2}.  We observe an upward trend in $C_{\rm o}\paren{\gamma}$ as a function of increasing values of $\gamma$, which is an expected result since small values of $\gamma$ correspond to more stringent outage constraints. Large values of $\kappa$ put a downward pressure on the $C_{\rm o}\paren{\gamma}$ due to increased levels of AWI. The last but not least, $C_{\rm o}\paren{\gamma}$ curves are almost perfectly tracked by our heuristic rate in almost all cases considered in Fig. \ref{fig2}.  

\section{Conclusions}\label{Conclusion}
In this paper, we have investigated the downlink outage performance of $K$-tier HCN networks under general signal propagation models, allowing for the use of general bounded path-loss functions and arbitrary fading distributions.  Tight upper and lower bounds on the outage probability and outage capacity have been obtained for two specific association policies - the generic association policy and the BARSS association policy.  The validity of our analytical results has also been confirmed by simulations.  The proposed approach can be extended to other association policies, and has the potential of understanding the HCN performance and design beyond specific selections of the path-loss model and the fading distribution.  

\vspace{-0.1cm}
%\section*{Acknowledgment}
%This research was supported in part by a Marie Curie FP7-Reintegration-Grants within the 7th European Community Framework Programme under Grant PCIG10-GA-2011-303713, in part by the Scientific and Technological Research Council of Turkey (TUBITAK) under Grant 115E162, and in part by the U.S. National Science Foundation under Grant ECCS-1343210.

%\vspace{-0.3cm}
\vspace{-0.1cm}
\appendices

\section{ The Proof of Theorem \ref{Thm: Outage Probability Bounds - Generic Association} } \label{Proof_Theorem_1}
In this appendix, we will provide the proof for Theorem \ref{Thm: Outage Probability Bounds - Generic Association} establishing the outage capacity bounds for the generic association policy. Given that the test user is associated to a BS at a distance $r$ in tier-$k$, we can express the $\tau$-outage probability as 
\begin{eqnarray*}
\PRP{\tau\mbox{-outage}}  &=& \PR{\log\paren{1 + \sinr_{\mathcal{A}} } < \tau}  \\
&=& \int_0^\infty \PR{I_{\vec{\lambda}} > P_k\paren{\frac{h G_k(r)}{\e{\tau} - 1} - \snr_k^{-1}}\PG} q_k(h) dh \\
&=& 1 - \int_{\frac{\snr_k^{-1}\paren{\e{\tau} - 1}}{G_k(r)}}^\infty \PR{I_{\vec{\lambda}} \leq P_k\paren{\frac{h G_k(r)}{\e{\tau} - 1} - \snr_k^{-1}}\PG} q_k(h) dh, 
\end{eqnarray*}
where the last inequality follows from the fact that $I_{\vec{\lambda}}$ is a positive random variable, and we have $P_k \paren{{\frac{{hG_{k}\left( r \right)}}{{{\e{\tau}} - 1}} - {\snr}_k^{ - 1}}} \PG < 0$ if and only if $h < \frac{{\left( {{\e{\tau} } - 1} \right){\snr}_k^{ - 1}}} {{G_{k}\left( r \right)}}$. By using Lemma \ref{Lemma: Gauss Approximation for Generic Policy} and the natural bounds $0$ and $1$ on the probability, we can upper and lower bound $\PRP{\tau\mbox{-outage}}$ as   
\begin{eqnarray*}
\PRP{\tau\mbox{-outage}} &\leq& 1 - \int_{\frac{\snr_k^{-1}\paren{\e{\tau} - 1}}{G_k(r)}}^\infty \max\brparen{0, \Psi\paren{\zeta_k\paren{h, \tau, r}} - \Xi \cdot c\paren{\zeta_k\paren{h, \tau, r}}} q_k(h) dh \\
&=& 1 - \ES{\max\brparen{0, \Psi\paren{\zeta_k\paren{H_k, \tau, r}} - \Xi \cdot c\paren{\zeta_k\paren{H_k, \tau, r}}} \I{H_k \geq \frac{\SNR_k^{-1}\paren{\e{\tau} -1}}{G_k(r)}}} \\
&=& 1 - \ES{V_k^{-}\paren{H_k, \tau, r}}
\end{eqnarray*}
and
\begin{eqnarray*}
\PRP{\tau\mbox{-outage}} &\geq& 1 - \int_{\frac{\snr_k^{-1}\paren{\e{\tau} - 1}}{G_k(r)}}^\infty \min\brparen{1, \Psi\paren{\zeta_k\paren{h, \tau, r}} + \Xi \cdot c\paren{\zeta_k\paren{h, \tau, r}}} q_k(h) dh \\
&=& 1 - \ES{\min\brparen{1, \Psi\paren{\zeta_k\paren{H_k, \tau, r}} + \Xi \cdot c\paren{\zeta_k\paren{H_k, \tau, r}}} \I{H_k \geq \frac{\SNR_k^{-1}\paren{\e{\tau} -1}}{G_k(r)}}} \\
&=& 1 - \ES{V_k^{+}\paren{H_k, \tau, r}},
\end{eqnarray*}
where $\Xi$ and $c(x)$ are as given in Lemma \ref{Lemma: Gauss Approximation for Generic Policy}, $\Psi(x)$ is the standard normal CDF and $\I{\cdot}$ is the indicator function.

\section{The Proof of Lemma \ref{Lemma: Association Probability} } \label{Proof_Lemma_3}
In this appendix, we will derive the connection probability of the test user to a serving BS in tier-$k$ under the BARSS association policy, which is denoted by $\PR{A^\star = k}$.  Let $R_i$ be the nearest distance from $\Phi_i$ to the test user for $i=1, \ldots, K$.  Then, utilizing the structure of the BARSS association policy, this probability can be written as
\begin{eqnarray}
\PR{A^\star = k} &=& \PRP{\bigcap_{ \myatop{i=1}{i \neq k} }^K \brparen{\beta_i P_i G_i\paren{R_i} \leq \beta_k P_k G_k\paren{R_k}}} \nonumber \\
&=& \int_0^\infty \PRP{\bigcap_{ \myatop{i=1}{i \neq k}  }^K \brparen{\beta_i P_i G_i\paren{R_i} \leq \beta_k P_k G_k\paren{u}} \bigg| R_k = u} f_{R_k}(u) du \nonumber \\
&\stackrel{\rm (a)}{=}& \int_0^\infty \prod_{ \myatop{i=1}{i \neq k} }^K \PR{\beta_i P_i G_i\paren{R_i} \leq \beta_k P_k G_k\paren{u} \Big| R_k = u} f_{R_k}(u) du \nonumber \\
&\stackrel{\rm (b)}{=}& \int_0^\infty \prod_{ \myatop{i=1}{i \neq k} }^K \PR{\beta_i P_i G_i\paren{R_i} \leq \beta_k P_k G_k\paren{u}} f_{R_k}(u) du, \label{Eqn: Connection Prob 1}
\end{eqnarray}
where the identity (a) follows from the conditional independence of the events 
$$\brparen{\beta_i P_i G_i\paren{R_i} \leq \beta_k P_k G_k\paren{R_k}} \hspace{0.15cm} \mbox{ for } \hspace{0.15cm} i \in \brparen{1, \ldots, K} \backslash \brparen{k} $$
given any particular realization of $R_k$, and the identity (b) follows from the independence of the nearest neighbour distances from different tiers.  Each probability term in \eqref{Eqn: Connection Prob 1} can be calculated as
\begin{eqnarray}
\PR{\beta_i P_i G_i\paren{R_i} \leq \beta_k P_k G_k\paren{u}} &=& \PR{R_i \geq G_i^{-1}\paren{\frac{\beta_k P_k}{\beta_i P_i} G_k(u)}} \nonumber \\
&=& \PR{R_i \geq Q_i^{(k)}(u)} \nonumber \\
&=& \exp\paren{-\pi \lambda_i \paren{Q_i^{(k)}(u)}^2}, \label{Eqn: Connection Prob 2}   
\end{eqnarray}
where the last equality follows from the nearest neighbour distance distribution for $R_i$.\footnote{The PDF and CDF of $R_i$ for $i=1, \ldots, K$ are given by $f_{R_i}(u) = 2\pi\lambda_i u \e{-\pi \lambda_i u^2}$ and $F_{R_i}(u) = 1 - \e{-\pi \lambda_i u^2}$, respectively.} Using \eqref{Eqn: Connection Prob 2}, we can write $\PR{A^\star = k}$ as

\begin{eqnarray}
\PR{A^\star = k} &=& \int_0^\infty \prod_{ \myatop{i=1}{i \neq k}  }^K \exp\paren{-\pi \lambda_i \paren{Q_i^{(k)}(u)}^2} f_{R_k}(u) du \nonumber \\ 
&=&  2 \pi \lambda_k \int_0^\infty u \exp\paren{-\pi \sum_{ \myatop{i=1}{i \neq k}  }^K \lambda_i \paren{Q_i^{(k)}(u)}^2}\exp\paren{-\pi \lambda_k u^2} du. \label{Eqn: Connection Prob 3}
\end{eqnarray}  

We note that some terms inside the summation $\sum_{i=1, i \neq k}^K \lambda_i \paren{Q_i^{(k)}(u)}^2$ may not be active for some particular values of $u$ if $G_k(u) \geq \frac{\beta_i P_i}{\beta_k P_k} G_i(0)$. Recalling the definition of $a_i \defeq \frac{\beta_i P_i}{\beta_k P_k} G_i(0)$, we observe that the condition $G_k(u) \geq \frac{\beta_i P_i}{\beta_k P_k} G_i(0)$ holds if and only if $u \leq G_k^{-1}\paren{a_i}$.  Introducing $a_0 = 0$ and $a_{K+1} = +\infty$ to have the integration limits from $0$ to $\infty$, and enumerating $a_i$'s in descending order for $i \neq k$, we finally arrive the desired result 
\begin{eqnarray}
\PR{A^\star = k} = 2\pi\lambda_k \sum_{j=1}^K \int_{r_{j-1}}^{r_j} u \exp\paren{-\pi\paren{\lambda_k u^2 + \sum_{i=1}^{j-1} \lambda_{\pi(i)} \paren{Q_{\pi(i)}^{(k)}(u)}^2}} du, \nonumber
\end{eqnarray}     
where $\pi(i)$ is an enumeration of $a_i$'s in descending order, i.e., $a_{\pi(i)} \geq a_{\pi(i+1)}$ for $i=0, \ldots, K-1$ and $r_i = G_k^{-1}\paren{a_{\pi(i)}}$ for $i=0, \ldots, K$.

\section{The Proof of Lemma \ref{Conditional_Connection_Distance}} \label{proof_Conditional_Connection_Distance}

In this appendix, we will derive the conditional PDF of the connection distance $R^\star$ given the event $\brparen{A^\star = k}$.  To this end, we will first calculate the conditional CDF of  $R^\star$ given $\brparen{A^\star = k}$, which will be denoted by $F_{R^\star | \brparen{A^\star = k}}(u)$.  Let $R_i$ be the nearest distance from $\Phi_i$ to the test user for $i=1, \ldots, K$.  Then,  
\begin{eqnarray}
F_{R^\star | \brparen{A^\star = k}}(u) &=& \PR{R^\star \leq u \big| A^\star = k} \nonumber \\
&=& \frac{1}{p_k^\star}\PR{R^\star \leq u \mbox{ and } A^\star = k} \nonumber \\
&=& \frac{1}{p_k^\star} \PR{R_k \leq u \mbox{ and } \bigcap_{ \myatop{i=1}{i \neq k} }^K \brparen{\beta_i P_i G_i(R_i) \leq \beta_k P_k G_k\paren{R_k}}}\nonumber \\
&=& \frac{1}{p_k^\star} \int_0^u \PR{\bigcap_{ \myatop{i=1}{i \neq k} }^K \brparen{\beta_i P_i G_i(R_i) \leq \beta_k P_k G_k\paren{r}} \bigg| R_k = r} f_{R_k}(r) dr.  \label{Eqn: Connection Distance 1}
\end{eqnarray} 

Using the conditional independence of the events $\brparen{\beta_i P_i G_i\paren{R_i} \leq \beta_k P_k G_k\paren{R_k}}$ for $i \in \brparen{1, \ldots, K} \backslash \brparen{k}$ for any given particular realization of $R_k$ and the independence of the nearest neighbour distances from different tiers, we can further simplify \eqref{Eqn: Connection Distance 1} as
\begin{eqnarray}
F_{R^\star | \brparen{A^\star = k}}(u) &=& \frac{1}{p_k^\star} \int_0^u \PR{\bigcap_{ \myatop{i=1}{i \neq k} }^K \brparen{\beta_i P_i G_i(R_i) \leq \beta_k P_k G_k\paren{r}} \bigg| R_k = r} f_{R_k}(r) dr \nonumber \\
&=&  \frac{1}{p_k^\star} \int_0^u \prod_{ \myatop{i=1}{i \neq k} }^K \PR{\beta_i P_i G_i\paren{R_i} \leq \beta_k P_k G_k(r)} f_{R_k}(r) dr \nonumber \\
&=& \frac{1}{p_k^\star} \int_0^u \prod_{ \myatop{i=1}{i \neq k} }^K \PR{R_i \geq G_i^{-1}\paren{\frac{\beta_k P_k}{\beta_i P_i}G_k(r)}} f_{R_k}(r) dr \nonumber \\
&=& \frac{1}{p_k^\star} \int_0^u \prod_{ \myatop{i=1}{i \neq k} }^K \exp\paren{-\pi \lambda_i \paren{Q_i^{(k)}(r)}^2} f_{R_k}(r) dr \nonumber \\
&=& \frac{2\pi\lambda_k}{p_k^\star} \int_0^u r \exp\paren{-\pi \paren{\lambda_k r^2 + \sum_{ \myatop{i=1}{i \neq k} }^K \lambda_i \paren{Q_i^{(k)}(r)}^2}}dr.  \label{Eqn: Connection Distance 2}
\end{eqnarray}

We obtain the conditional PDF of $R^\star$ given $A^\star = k$ by differentiating \eqref{Eqn: Connection Distance 2} with respect to $u$. This calculation leads to 
\begin{eqnarray}
f_k(u) = \frac{2\pi\lambda_k}{p_k^\star} u \exp\paren{-\pi \paren{\lambda_k u^2 + \sum_{ \myatop{i=1}{i \neq k} }^K \lambda_i \paren{Q_i^{(k)}(u)}^2}} \hspace{0.25cm} \mbox{ for } u \geq 0. \label{Eqn: Connection Distance 3}
\end{eqnarray}
We observe that the summation term appearing in \eqref{Eqn: Connection Distance 3} is exactly the same one appeared in \eqref{Eqn: Connection Prob 3}. Hence, the same enumeration step can be carried out to arrive at the final result 
\begin{eqnarray}
f_k(u) = \frac{2\pi\lambda_k}{p_k^\star} \sum_{j=1}^K u \exp\paren{-\pi\paren{\lambda_k u^2 + \sum_{i=1}^{j-1} \lambda_{\pi(i)} \paren{Q_{\pi(i)}^{(k)}(u)}^2}} \I{u \in [r_{j-1}, r_j)}, \nonumber
\end{eqnarray}
where $a_0 = 0$, $a_{K+1} = +\infty$, $a_i = \frac{\beta_i P_i}{\beta_k P_k} G_i(0)$ for $i \in \brparen{1, \ldots, K} \backslash \brparen{k}$, $\pi(i)$ is an enumeration of $a_i$'s in descending order, i.e., $a_{\pi(i)} \geq a_{\pi(i+1)}$ for $i=0, \ldots, K-1$ and $r_i = G_k^{-1}\paren{a_{\pi(i)}}$ for $i=0, \ldots, K$.

\ifCLASSOPTIONcaptionsoff
  \newpage
\fi



%\bibliographystyle{IEEEtran}

%\begin{biography}[{\includegraphics[width=1in,height=1.25in,clip,keepaspectratio]{mshell}}]{Michael Shell}

\vfill \vfill
\end{document}